\documentclass[proceedings]{rmaa}

\usepackage{rmaacite}

\renewcommand{\P}[1]{%
\ifnum#1=1\hbox{OW~168--326E}\fi
\ifnum#1=2\hbox{OW~167--317}\fi
\ifnum#1=3\hbox{OW~163--317}\fi
\ifnum#1=5\hbox{OW~158--323}\fi
\ifnum#1=0\hbox{OW~171--334}\fi}

\title{Introducing ZEUS-MP: A 3D, Parallel, Multiphysics Code
for Astrophysical Fluid Dynamics}
\author{Michael L. Norman\altaffilmark{1}
  \affil{Astronomy Department and NCSA, University of Illinois, Urbana-Champaign} }
\altaffiltext{1}{Laboratory for Computational Astrophysics}           

\fulladdresses{
\item Michael L. Norman, Beckman Institute, 405 N. Mathews St.,
  Urbana, IL, 61801 (norman@ncsa.uiuc.edu)} 

\shortauthor{Norman}
\shorttitle{Introducing ZEUS-MP}                     


\abstract{  We describe ZEUS-MP: a Multi-Physics, Massively-Parallel, Message-Passing code for
astrophysical fluid dynamics simulations in 3 dimensions. ZEUS-MP is a follow-on to the sequential
ZEUS-2D and ZEUS-3D codes developed and disseminated by the Laboratory for Computational 
Astrophysics ({\tt lca.ncsa.uiuc.edu}) at NCSA. 
V1.0 released 1/1/2000 includes the following physics
modules: ideal hydrodynamics, ideal MHD, and self-gravity. Future releases will include
flux-limited radiation diffusion, thermal heat conduction, two-temperature plasma, and
heating and cooling functions. The covariant equations are cast on a moving Eulerian grid with
Cartesian, cylindrical, and spherical polar coordinates currently supported. Parallelization is done
by domain decomposition and implemented in F77 and MPI. The code is portable across a wide
range of platforms from networks of workstations to massively parallel processors.
Some parallel performance results are presented as well as an
application to turbulent star formation.
}

\nonstopmode

\begin{document}

\maketitle

\section{A Brief History of ZEUS}
\label{sec:intro}
ZEUS is a family of codes for astrophysical fluid dynamics simulations
developed at the Laboratory for Computational Astrophysics (LCA)
of the National Center for Supercomputing Applications (NCSA) at the University
of Illinois, Urbana-Champaign. The purpose of this paper is to announce the availability
of the latest implementation: ZEUS-MP. Version 1.0 released 1/1/2000 is available from
the LCA's website {\tt lca.ncsa.uiuc.edu}.

ZEUS has its roots in a 2D Eulerian hydro code developed by M. Norman for simulations of
rotating protostellar collapse \cite{NWB80} while he was a student at the
Lawrence Livermore National Laboratory. The hydrodynamics algorithm, which has changed little
in subsequent versions, is based on a simple staggered-grid finite-difference scheme
\cite{N80,NW86}.
Shock waves were captured within a few cells with a von Neumann-Richtmyer type artificial viscosity.
A powerful and essential feature of the code, which has been retained in subsequent versions,
was that the equations of self-gravitating hydrodynamics were solved on a moving Eulerian grid
permitting accurate simulation over a range of scales in the collapsing protostar.

A significant improvement to the hydrodynamics algorithm came with the incorporation of the
second order-accurate, monotonic advection scheme \cite {VL77}.
This code version, innocuously called A2, was vectorized for the Cray-1 supercomputer
at the Max-Planck-Institut f\"ur Astrophysik, and extensively applied to the simulation
of extragalactic radio jets \cite{NSWS82}.

The first code called ZEUS was developed by David Clarke as a part of his PhD thesis 
on MHD jets \cite{Clarke88,CNB86} under Norman's supervision. One of the principal
challenges in numerical MHD simulations is satisfying the zero-divergence constraint on $B$.
In our axisymmetric simulations, this was insured by evolving the toroidal component of the 
magnetic vector potential from which divergence-free poloidal field components can be derived, as
well as evolving the toroidal magnetic field component directly. Third order-accurate monotonic
advection was used for evolving $A_{\phi}$ in order to improve the quality of the derived
current densities.

The next development was a major rewrite and significant extension of ZEUS
by James Stone as a part of his PhD thesis at the University of Illinois. The resulting
code, named ZEUS-2D, solves the equations of self-gravitating radiation magnetohydrodynamics in
2D or 2-1/2D. Many new algorithms were developed and incorporated into ZEUS-2D including:
(1) a covariant formulation, allowing simulations in various coordinate geometries; (2) a tensor artificial
viscosity; (3) a new, more accurate MHD algorithm (MOC-CT) combining the Constrained Transport
algorithm \cite{EH88} with a Method Of Characteristics treatment for Alfv\'en waves; and (4)
a variable tensor Eddington factor solution for the equations of radiation hydrodynamics.
ZEUS-2D's algorithms and tests are described in detail in a series of three papers 
\cite{SN92a,SN92b,SMN92}(the ZEUS Trilogy).

The MOC-CT algorithm for numerical MHD was specifically designed to be extensible to 3D, and work
on a 3D version of ZEUS began in 1989 when David Clarke came to Illinois as Norman's postdoc.
Written for the Cray-2 supercomputer, ZEUS-3D physics options included hydrodynamics, MHD,
self-gravity, and optically thin radiative cooling. Parallelization was done using Cray
Autotasking compiler directives. Novel features of the code included the use of a custom 
source code pre-processor which handled a variety of source code transformations. 
Another useful feature of ZEUS-3D was an extensive set of inline
graphics and diagnostic routines, as well as the ability to run in 1D and 2D mode.   

With a grant from the National Science Foundation in 1992, the LCA was established with 
the purpose of disseminating ZEUS-2D, ZEUS-3D and the TITAN implicit adaptive-mesh radiation
hydrodynamics code \cite{GM94} to the international
community. Currently, there are over 500 registered users of LCA codes in over 30 countries.
Some recent applications of ZEUS include planetary nebulae \cite{GGS99}, molecular
cloud turbulence \cite{MML99}, and solar magnetic arcades \cite{LM00}. 

Work was begun on ZEUS-MP in the fall of 1996 by Robert Fiedler and subsequently
by John Hayes and James Bordner with support from the Department of Energy to explore 
algorithms for parallel radiation hydrodynamics simulations in 3D.     

\section{Why ZEUS-MP?}

ZEUS-MP is a portable, parallel rewrite of ZEUS-3D. MP stands for: {\em Multi-Physics},
{\em Massively-Parallel}, and {\em Message-Passing}. 3D simulations are by their nature
memory-- and compute--intensive. The most powerful computers available today 
are parallel computers with hundreds to thousands of processors connected into
a cluster. While some systems offer a shared memory view to the applications programmer,
others, such as Beowulf clusters, do not. Thus, for portability sake we have
assumed ``shared nothing" and implemented ZEUS-MP as a SPMD (Single Program, Multiple Data)
parallel code using the MPI message-passing library to affect interprocessor
communication.

\begin{figure}
  \begin{center}
    \leavevmode
    \includegraphics[width=0.8\textwidth]{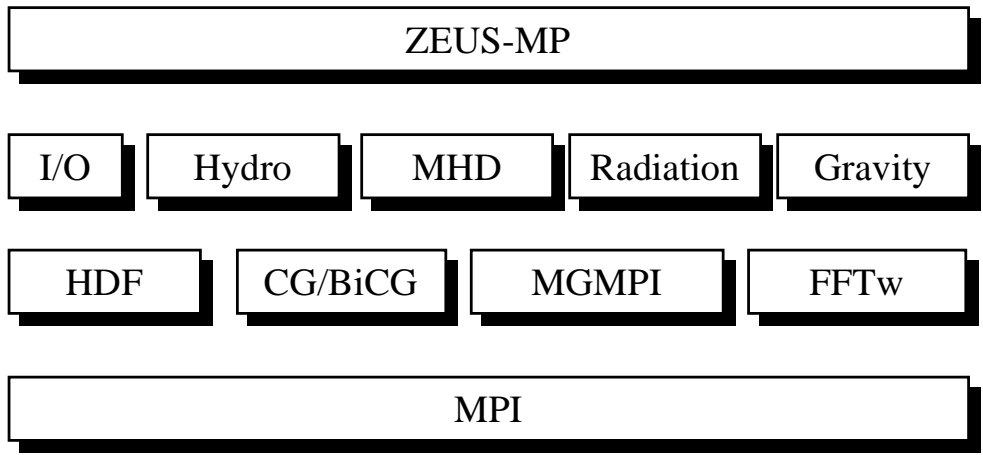}
    \caption{Block diagram of the ZEUS-MP code.} 
   \label{fig:diagram}
  \end{center}
\end{figure}

Figure 1 shows a block diagram of the major components of ZEUS-MP.
ZEUS-MP is composed of an application layer (second row) and a
libraries layer (third row), all resting on the MPI message
passing library. A brief description follows. 

There are four main physics modules: hydro, MHD, radiation
transport, and self-gravity. The hydro and MHD modules are
time-explicit, and thus require no linear algebra libraries
for their solution. The hydro algorithm is a straight-forward
3D extension of the algorithm described in \cite{SN92a}. 
The MHD algorithm is the MOC-CT
algorithm described in \cite{SN92b} with the modifications
described in \cite{HS95} for enhanced stability in weakly
magnetized, strongly sheared flows. At present, nonideal
effects (viscosity, resistivity) are not included, however
one can choose between an ideal gamma-law or isothermal
equations of state. The radiation transport algorithm   
implements the time-implicit flux-limited diffusion
algorithm of Stone (1994). The radiation and gas
energy equations are solved as a coupled, implicit system,
resulting in a large, sparse, banded system of linear
equations which must be solved within an outer nonlinear Newton
iteration. Two linear system solvers are built into ZEUS-MP:
a conjugate gradient solver (CG/BiCG) with
diagonal preconditioning, and a multigrid solver (MGMPI)
\cite{Bordner00}. Problems involving self-gravity require
the solution of the Poisson eqation. Two Poisson solvers
are built into ZEUS-MP: a Fourier space solver for
triply periodic cubic grids, and an elliptic finite
difference solver for all other geometries and boundary
conditions. The former utilizes the FFTw library developed
at MIT, while the latter uses MGMPI.  

As in earlier versions of ZEUS, the equations solved by
ZEUS-MP are formulated on a covariant, moving Eulerian
grid. Problems in Cartesian, cylindrical, and spherical
polar coordinates can be run with a variety of boundary
conditions and types (periodic, Dirichlet, Neumann). 
The linear system solvers are designed
to handle all cases.

File I/O is done using NCSA's HDF (Hierarchical Data File)
standard \cite{Folk00}, which is a widely adopted portable
file format for scientific data. 

\section{Parallelism}

\begin{figure}
  \begin{center}
    \leavevmode
    \includegraphics[width=0.6\textwidth]{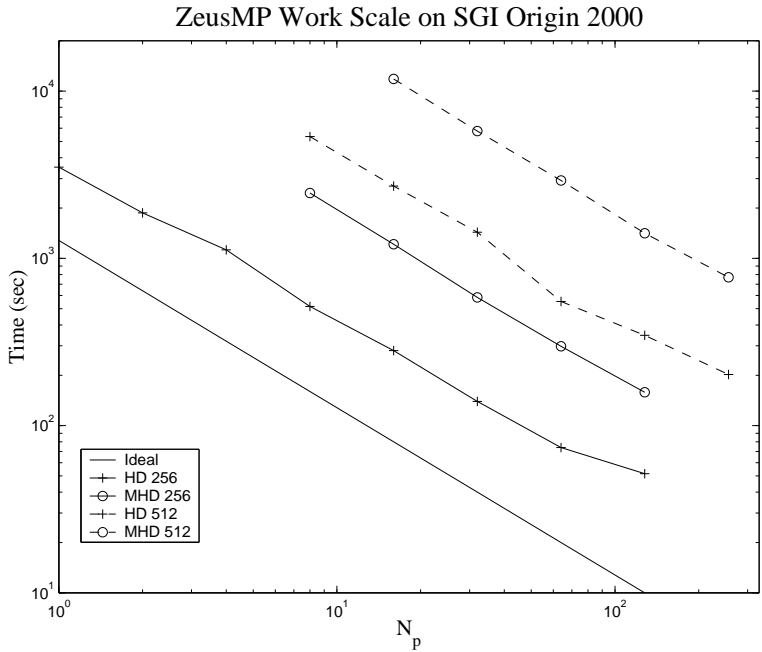}
    \caption{Execution time for ZEUS-MP parallel
     benchmarks on the SGI/Cray Origin2000. Curves
     plot time for 30 timesteps for ``fixed work"
     scaling tests for hydrodynamic
     and MHD blast tests at $256^3$ and $512^3$ 
     resolution. The solid line without symbols
     shows ideal linear speedup. As can be seen, ZEUS-MP
     achieves near-ideal scaling.} 
   \label{fig:scaling}
  \end{center}
\end{figure}

ZEUS-MP utilizes {\em domain decomposition} \cite{Foster95}
for parallelization, wherein the computational domain is
subdivided into a number of equally-sized regions, each of
which is assigned to a different processor for execution. 
Depending upon the problem size and the number of processors
targeted, the user can specify a 1D ``slab", 2D ``pencil",
or 3D ``block" decomposition. A region is represented in
processor memory as arrays
of data storing the solution vector for a specific subdomain.
The arrays are dimensioned so as to include two layers of 
buffer zones on each face of the block for the purpose of transferring
boundary conditions from neighboring processors. Data
transfer between neighboring blocks, as well as collective
operations and global reductions are handled via MPI function
calls. 

ZEUS-MP performance has been optimized in two ways \cite{Fiedler97}: 
(1) single processor performance, using a variety of standard
cache optimization techniques; and (2) parallel performance,
using asynchronous communication, wherein computation and
communication is overlapped. Single-node performance on an
MIPS R10000 processor is in the range of 100 MFlop/s. Parallel 
speedup results depend sensitively on the properties of the
network hardware and software on the host computer. A
collection of benchmark results can be found on the ZEUS-MP
website {\tt zeus.ncsa.uiuc.edu/lca\_intro\_zeusmp.html}. 
Figure 2 shows near-ideal scaling on an SGI/Cray Origin2000 for
hydrodynamic and MHD tests of size $256^3$ and $512^3$
on up to 256 processors.   

\section{Source Code Availability}

V1.0 is available now as a downloadable tar file from the ZEUS-MP
website. Rudimentary
documentation describing how to make the executable and run
any of the five test problems is also available online.
At present, the radiation module is not included; pending
clarification of export restrictions, it will be made available
in a later release.  

Also available is a new 3D visualization tool called LCA Vision
{\tt zeus.ncsa.uiuc.edu/$\sim$miksa/LCAVision1.0.html}. Vision reads
HDF files and provides a variety of visualization tools in an 
easy-to-use, menu-driven interface.

\section{Sample Application}

\begin{figure}
  \begin{center}
    \leavevmode
    \includegraphics[width=0.5\textwidth]{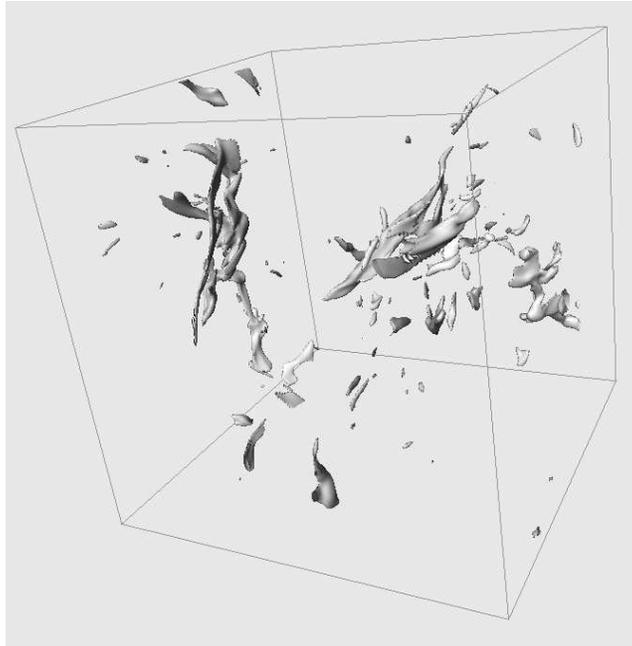}
    \caption{ZEUS-MP simulation of a turbulent, magnetized, self-gravitating
             molecular cloud with $M/M_{cr}=1.1$. An isosurface at 8 times the 
             mean density is shown. The mean magnetic field is parallel to the Z-axis.
             Gravitational collapse of filamentary clouds is found, despite 
             supersonic (M=5) driving.} 
   \label{fig:turbulence}
  \end{center}
\end{figure}

To illustrate ZEUS-MP's capabilities, we present an application to 
magnetic star formation.  
The purpose is to understand the competition between turbulent,
magnetic, and gravitational stresses in the formation of gravitationally 
bound cores in a turbulent molecular cloud \cite{HMK00}. 
In particular, we want to explore whether the critical mass-to-flux
ratio \cite{MS76} is a good predictor of gravitational collapse in
a cloud driven by turbulence.

The calculation is done in a triply periodic cube with $256^3$ cells
distributed across 64 processors ($4^3$ cubic decomposition.)
One begins with a uniform density gas filling the box threaded by a uniform
magnetic field in the Z-direction. The gas is assumed to be isothermal.
A turbulent velocity field is established by driving in Fourier space
over a limited range of wavenumbers as described in \cite{MML99}. 
A statistical steady state is reached within a few dynamical times,
and once it has, gravity is switched on. Thereafter, density peaks formed by
colliding gas streams may become gravitationally bound and collapse. 
The problem is described by a few dimensionless parameters formed by
combinations of
the gas density, sound speed, magnetic field strength, rms velocity 
perturbation,  and box length,: $n_J=12$, the number of thermal Jeans' masses
in the box; $M/M_{cr}=1.1$, the ratio of the box mass to the critical mass
that can be supported by a static magnetic field; $M_{turb}=5$, the
turbulent Mach number; and $k_{drv}$=2, the driving wavenumber in units 
of the inverse box length. 

Fig. 3 shows isosurfaces of gas density at 1.82 freefall times after
gravity was swithched on. The isolevel is eight times the initial
uniform density. One can see evidence
for filamentary and flattened condensations which are both aligned 
and perpendicular to the mean field direction.  
Gravitational collapse is already underway, as indicated by high densities
in the centers of several filaments. We find, as did \cite{HMK00} using
ZEUS-3D, that supersonic turbulence is not able to prevent gravitational 
collapse from occurring in magnetically supercritical clouds. Simulations
at $512^3$ grid resolution and beyond are underway to investigate this
further.

\acknowledgements I gratefully acknowledge my colleagues, past and present,
who have materially contributed to ZEUS-MP: Jim Stone and David Clarke for
their foundational contributions embodied in ZEUS-2D and ZEUS-3D; 
Robert Fiedler, who wrote the first hydrodynamics version of ZEUS-MP code and
optimized it for cache-based parallel systems; John ``radiation cowboy" Hayes, 
who implemented the implicit radiation diffusion module along with the CG/BiCG linear solver;
Mordecai-Mark MacLow, who ported the HSMOC MHD algorithm from ZEUS-3D, which
in turn was incorporated by Byung-Il Jun; Pakshing Li,
who developed and tested both Poisson solvers and prepared the V1.0 release; 
and James Bordner, developer of the MGMPI multigrid solver
and patient fixer of what we break. I thank my collaborators Fabian Heitsch,
Mordecai MacLow and Pakshing Li for allowing me to show our unpublished results
(Fig. 3). This work has been supported by contracts
B324163 and B506131 from the Lawrence Livermore National Laboratory. I would
like to thank Frank Graziani of B Division for his continued interest and support.


\end{document}